# Development and characterization of a 2.2 W narrow-linewidth 318.6 nm ultraviolet laser

JIEYING WANG [1, 2], JIANDONG BAI [1, 2], JUN HE [1, 2, 3], AND JUNMIN WANG [1, 2, 3,*]

[1]State Key Laboratory of Quantum Optics and Quantum Optics Devices, Shanxi University, Tai Yuan 030006, Shan Xi Province, China
[2]Institute of Opto-Electronics, Shanxi University, Tai Yuan 030006, Shan Xi Province,
[3]Collaborative Innovation Center of Extreme Optics, Shanxi University, Tai Yuan 030006, Shan Xi Province, China
* Corresponding author: wwjjmm@sxu.edu.cn

We demonstrate a high-power narrow-linewidth ultraviolet (UV) laser system at 318.6 nm for direct $6S_{1/2}$-nP (n = 70 to 100) Rydberg excitation of cesium atoms. Based on commercial fiber lasers and efficient nonlinear frequency conversion technology, 2.26 W of tunable UV laser power is obtained from cavity-enhanced second harmonic generation following sum-frequency generation of two infrared lasers at 1560.5 nm and 1076.9 nm to 637.2 nm. The maximum doubling efficiency is 57.3%. The typical UV laser power root-mean-square fluctuation is less than 0.87% over 30 minutes, and the continuously tunable range of the UV laser frequency is more than 6 GHz. Its beam quality factors $M_X^2$ and $M_Y^2$ are 1.16 and 1.48, respectively. This high-performance UV laser has significant potential use in quantum optics and cold atom physics.

## 1. INTRODUCTION

Highly excited Rydberg atoms have played an important role in quantum information processing and quantum optics. The long lifetime, large dipole matrix element and strong interaction between neighboring atoms make the Rydberg blockade mechanism ideal for quantum memory [1, 2] and quantum computing [3]. In particular, the strong interaction enables it to become a promising candidate to observe entanglement [4], many-body Rabi oscillations [5], and to implement elementary quantum gates [6]. In most of these experiments, given available lasers in the visible or near-IR, most often a two-step or three-step cascaded excitation is chosen to produce a desired Rydberg state. However, in these techniques photon scattering and ac Stark shifts from the intermediate state introduce decoherence, dipole forces, and frequency noise. Minimizing photon scattering is of great significance when using Rydberg-dressed atoms to create tunable, long lived, many-body interactions in a quantum gas [7]. At present, studies of Rydberg excitation with a single-photon transition are rare, first demonstrated by Tong *et al.* [8] in 2004 in Rb at 297 nm and by Hankin *et al.* [9] in 2014 in Cs at 319 nm.

While avoiding the disadvantage of cascaded ladder-type excitation, the most challenging aspect of the single-photon technique is the requirement of a high-power, narrow-linewidth UV laser source tuned to the transition. Usually UV lasers are attained by nonlinear frequency conversion. However, this is limited by the UV absorption in the nonlinear materials and the associated thermal effects. Recently developed nonlinear crystals and low-loss coatings have allowed frequency doubling of the widely tunable Ti:sapphire laser and liquid dye lasers to achieve UV laser output [10-12]. However, such systems are large and expensive, and their operation and maintenance are very complex. Newly-developed fiber lasers and amplifiers are more simple and reliable to operate and maintain [13, 14]. For years, a number of nonlinear materials have been used to generate the UV, and among them $\beta$-BaB$_2$O$_4$ (BBO) and LiB$_3$O$_5$ (LBO) are prime choices. Both have very high optical damage thresholds. Their phase matching can be achieved by crystal temperature and orientation tuning. Angle-tuned phase matching is preferable in practice because of the high temperatures required for phase matching at UV wavelengths. LBO has a smaller birefringent walk-off angle, but BBO has a larger effective nonlinear coefficient ($d_{eff}$ of nearly 2 pm/V versus approximately 0.75 pm/V for LBO) [15] and a wide phase matched transparency down to 205 nm. Generally BBO is chosen for doubling to around 319 nm and it has experimentally demonstrated better performance [9, 16, 17]. In 2011, Wilson *et al.* first achieved UV laser output from two infrared fiber laser sources, with 750 mW of 313 nm laser emission produced for the laser cooling of Be$^+$ ions [16]. In 2014, for the same objective, 1.9 W of 313 nm UV laser output was produced by Lo *et al.* [17]. In 2014, Hankin *et al.* achieved 300 mW at 319 nm and were the first to use it for the single-photon Rydberg excitation of cesium [9]. For the same end purpose, we intend to produce high-power, stable and reliable 318.6-nm laser emission. This paper discusses our progress and accomplishments.

Previously [18] we detailed our experiments in sum-frequency generation (SFG). Using a single pass SFG configuration, we were able to obtain 8.75 W at 637.2 nm by adding two infrared beams at 1560.5 nm and 1076.9 nm in a 40 mm length of MgO doped periodically poled lithium niobate (PPMgO:LN). In the present work, taking the 637.2 nm red light we perform a thorough study on efficient second harmonic generation (SHG) with a BBO crystal in a ring cavity to create tunable 318.6 nm UV. In our scheme, we introduce the modulation needed for Pound–Drever–Hall (PDH) locking [19] of the ring cavity into the 1560.5-nm seed laser prior to initial amplification and SFG process so that a conventional electro-optic phase modulator (EOPM) can be used without degradation. Using an optimally-designed bow tie ring cavity, we demonstrate 2.26 W UV laser output, with a maximum conversion efficiency of 57.3%. The power stability, beam profile, and the continuously tunable range are discussed. The linewidth is estimated to be less than 10 kHz. This 318.6 nm laser will directly excite $6S_{1/2}$ ground state cesium atoms to nP (n = 70 to 100) Rydberg levels.

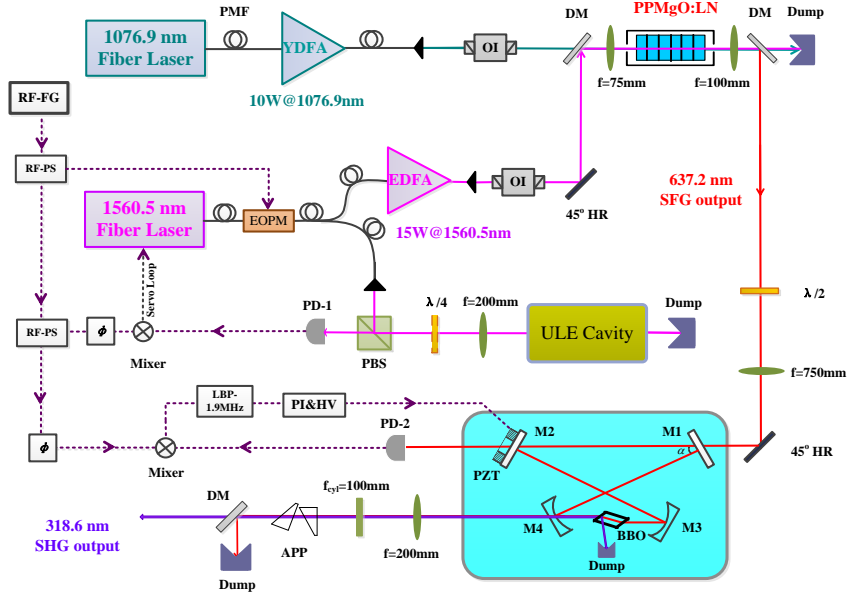

**Fig. 1.** Schematic diagram of the 318.6 nm UV laser system. Two infrared lasers are frequency summed in a PPMgO:LN crystal to generate 637.2 nm red light, which is then frequency doubled to 318.6 nm in a BBO crystal via a symmetric bow tie ring cavity. The cavity is actively stabilized by the PDH method. Keys to the figure: EDFA: erbium doped fiber amplifier; YDFA: ytterbium doped fiber amplifier; PMF: polarization maintaining optical fiber (all fiber loops); OI: optical isolator; λ/2: half-wave plate; PBS: polarization beam splitter cube; λ/4: quarter-wave plate; DM: dichroic mirror; 45° HR: 45° high reflectivity mirror; ULE cavity: ultralow expansion cavity; M1, M2, M3, M4: cavity mirrors; APP: anamorphic prism pair; EOPM: electro-optic phase modulator; PD: photodiode; PZT: piezoelectric transducer; RF-FG: radio frequency function generator; RF-PS: radio frequency power splitter; φ: phase shifter; LBP: low band pass filter; PI: proportional integration amplifier; HV: high voltage amplifier.

## 2. EXPERIMENTAL SETUP

The experiment setup is shown in Fig. 1. We demonstrate the 318.6-nm UV laser by cavity enhanced SHG of 637.2 nm red light which is produced from SFG of two infrared lasers. We start with two distributed feedback seed fiber lasers at 1560.5 nm and 1076.9 nm, with linewidths of approximately 600 Hz and 2 kHz [20], fed through separate doped fiber amplifiers. The 1560.5-nm seed laser is phase modulated. Its output is coupled into a waveguide-type EOPM (EO Space model PM-0S5-10-PFA-PFA-UL) running at 12.6 MHz and is split into two fibers. One is coupled into the fiber amplifier for locking the ring cavity using the PDH sideband modulation technique. The second is injected into an ultralow expansion (ULE) cavity, and then is fed back via a servo loop to the piezoelectric transducer (PZT) of the 1560.5 nm seed fiber laser for frequency stabilization. A 40 mm PPMgO:LN crystal is chosen for SFG to produce the 637.2 nm red light. The single frequency red light passes through a half-wave plate to adjust the polarization, and then is coupled into the ring cavity by a mode-matching 750 mm focal length lens.

The self-designed ring cavity is in a symmetric bow tie with two plane mirrors M1 and M2 and two plano-concave mirrors M3 and M4 of 100 mm radii of curvature. M1 is the input coupler with a transmission $T_1$ of 2.2%, and back faces antireflection (AR) coated at 637.2 nm. $T_1$ is optimized for impedance matching which allows the maximum coupling efficiency of the incident power. The other cavity mirrors are high reflectivity mirrors for p-polarized 637.2 nm laser ($R_p > 99.9\%$), and the output mirror M4 has a 94.5% transmission at 318.6 nm. The back faces of the high reflectors are uncoated. M2 is mounted on a PZT to lock the ring cavity on the resonance with the 637.2 nm light via the PDH method. A BBO crystal (Castech Inc.) 10 mm in length together with a laboratory-constructed copper oven mounted on a four-axis tilt aligner (New Focus model 9071) are placed at the waist between the plano-concave mirrors. In caution over possible high UV power damage, we use a Brewster cut BBO crystal rather than an AR-coated square cut crystal. The Brewster angle is 59.1°, the phase-matching angle $\theta$ is 37.6°, and the azimuthal angle $\varphi$ is 0°. It is cut for type I phase matching near room temperature, and its phase-matching condition is realized by orientation tuning. BBO crystal is highly hydroscopic with largest storage humidity of 30%. In order to achieve long life of the crystal, the entire cavity is housed in a relatively sealed acrylic box with silica gel to reduce humidity.

The ring cavity is designed by determining the folding angle and mirror separations to achieve an optimum waist within the BBO crystal. According to Boyd and Kleinman theory [21], the optimum confocal condition for an incident Gaussian beam and the desired beam waist is related to the crystal length. Optimal SHG occurs when the focusing parameter $\xi = l/b = 2.84$, where $l$ is the length of the BBO crystal, $b = 2\pi w_0^2/\lambda$, and $w_0$ is the beam waist. This condition results in an optimized waist of 20.9 μm in our case. However, as the power of the 637.2 nm laser reach 4 W, the circulating power in the cavity reaches nearly 200 W. To circumvent thermal effects and possible damage to the crystal, we modestly ease the confocal focusing in the BBO crystal [22]. By numerical simulation, a set of newly optimized cavity parameters gives a total length of 616 mm, with 116 mm between M3 and M4. Taking into consideration astigmatism compensation, the folding angle α is 10.7° (see Fig. 1). In this configuration, the 637.2-nm horizontal and vertical beam waist radii within the BBO crystal are 36.3 μm and 34.9 μm, and the second waists between the plane mirrors are 199.3 μm and 199.8 μm. The wider focusing in the crystal can nonetheless achieve considerable conversion efficiency. The maximum UV laser output at a given incident power can be achieved by carefully adjusting the phase-matching angle of the BBO crystal. The generated UV beam transmits through M4 as shown in Fig. 1, and then is separated from the non-converted red beam with dichroic mirrors after the ring cavity.

## 3. EXPERIMENTAL RESULTS AND DISCUSSION

The circulating power $P_c$ in the ring cavity in relation to the incident power $P_{in}$ is given by the iterative relation [23],

$$P_c = \frac{T_1 \cdot P_{in}}{\left[1-\sqrt{(1-T_1)(1-L)(1-E_{nl} \cdot P_c)}\right]^2} \quad (1)$$

where $T_1$ is the transmission of the input coupler, $L$ is the linear loss in the cavity, and $E_{nl}$ is the conversion coefficient of the BBO crystal. The 318.6 nm second harmonic power $P_{SHG}$ depends quadratically on $P_C$ by

$$P_{SHG} = E_{nl} \frac{(T_1 \cdot P_{in})^2}{\left[1-\sqrt{(1-T_1)(1-L)(1-E_{nl} \cdot P_c)}\right]^4} \quad (2)$$

The ratio of $P_{SHG}$ to $P_C$ gives the doubling efficiency $\eta$,

$$\eta = \frac{E_{nl} \cdot P_{in} \cdot T_1^2}{\left[1-\sqrt{(1-T_1)(1-L)(1-E_{nl} \cdot P_c)}\right]^4} \quad (3)$$

To achieve a high doubling efficiency at a given power value, optical impendence matching must be taken into account. The optimum transmission $T_{opt}$ of the input coupler is related to the overall losses including linear (loss of all elements in the cavity) and nonlinear losses (power converted to SHG). It can be written [24] as:

$$T_{opt} = \frac{L}{2} + \sqrt{\left(\frac{L}{2}\right)^2 + E_{nl} \cdot P_{in}} \quad (4)$$

The single-pass conversion coefficient $E_{nl}$ is measured by removing the input coupler M1, which preserves the focusing condition used in the actual resonant cavity. The measured $E_{nl}$ is $6.5 \times 10^{-5}$ /W from readings at different incident 637.2 nm laser powers. The total linear losses are found by measuring the cavity finesse on replacement of M1 with a high reflection plane mirror. The finesse is 937 at an input power of 3 mW, kept low so that the nonlinear conversion loss is in the $10^{-4}$ range and could be ignored. The overall linear loss $L$ is 0.67%. With these values of $E_{nl}$ and $L$ the optimum input coupler transmission at $P_{in}$ of 4 W is 2.0%. We select a transmission of 2.2% for the input coupler, which is very close to $T_{opt}$.

To lock the ring cavity, we modulated the 1560.5-nm seed laser at a frequency Ω of 12.6 MHz, thus avoiding the use of a high-power EOPM with attendant concern over thermal instability. Because the fiber amplifier has a broad gain bandwidth and energy is conserved in the SFG process, the sum frequency at 637.2 nm will be efficiently modulated (Fig. 2) and can be used to actively stabilize the ring cavity. Another advantage of this methodology is that the same EOPM can be used to lock the 1560.5-nm seed laser to a high-finesse ULE cavity via the PDH method. One disadvantage is that the modulation sideband is present in the UV output. A 318.6-nm Fabry–Pérot cavity analyzes the UV laser power in the sidebands. Because the modulation frequency is about six times the doubling cavity linewidth of approximately 2.2 MHz, the power of sidebands are sharply suppressed following SHG by about three orders of magnitude relative to the carrier frequency. It can be ignored in most cases. Should this UV laser be used in spectroscopy, the modulation frequency can be tuned to avoid excitation of any unwanted transition. In addition, for high-resolution spectroscopy of cold cesium Rydberg atoms, we have calculated that at principal quantum numbers in the range 70 to 100 the energy level spacing is around 10 GHz, and the splitting between the fine energy level $P_{1/2}$ and $P_{3/2}$ is around 400 MHz, so the weak UV sidebands can't excite unwanted transitions.

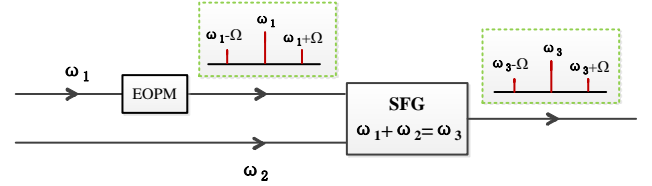

**Fig. 2.** Modulation sideband transfer in the SFG process. The 1560.5-nm seed laser is at $\omega_1$, and Ω is the 12.6 MHz modulation frequency applied to the EOPM. This generates sidebands which are amplified faithfully by the fiber amplifier because of its wide bandwidth. The 1076.9 nm laser at $\omega_2$ is not modulated. In the SFG process, the output red laser at $\omega_3$ contains the modulation.

After careful mode matching, impedance matching and adjustment of the phase-matching angle of the BBO crystal, the measured output power of the 318.6-nm laser and the corresponding doubling efficiency are shown in Fig. 3. The power is measured after a collimating lens and three dichroic mirrors with a total transmission of 92%. The experimental data are shown as purple and blue symbols, and the solid curves are the theoretical calculations from equations (2) and (3) with $T_1$ = 2.2%, $L$ = 0.67%, and $E_{nl}$ = 6.5×10$^{-5}$ /W. At an input power of over 2 W, the conversion efficiency begins to saturate at about 56%. The maximum UV output power is 2.26 W. The maximum doubling efficiency $\eta_{max}$ is 57.3% for an incident power of 3.30 W. Since the SHG process satisfies type I phase matching, the polarization of the 318.6-nm beam is orthogonal to the 637.2-nm beam, and the Brewster-cut crystal surface introduces a Fresnel reflection loss of 16.5% for the s-polarized UV beam. Accounting for this loss, the maximum internal efficiency is 68.6%. We find that the maximum doubling efficiency occurs at a BBO crystal temperature of 19 °C. As shown in Fig. 3, the calculated power and efficiency curves agree well with the experimental data at low input power levels. However, they diverge at high power levels. This is primarily caused by a decrease of the cavity coupling efficiency [25]. We have discovered that when the 637.2-nm red laser is mode matched into the bow tie ring cavity with a 750-mm focal length lens, the typical mode-matching efficiency at low power is about 98%, while at a power of 4 W it is only 80%. This mainly arises from thermal lensing in the PPMgO:LN crystal in the high-power regime of the SFG process [23]. This degrades the spatial mode of the 637.2-nm beam, leading to a low mode-matching efficiency. This could be improved by using a fiber to filter the mode of the red laser in front of the cavity or repeating the mode matching at the high incident power point. Additionally, absorption and thermal effects in the BBO crystal are no longer negligible at Watt-level UV power, and this adds the discrepancy between theory and experiment.

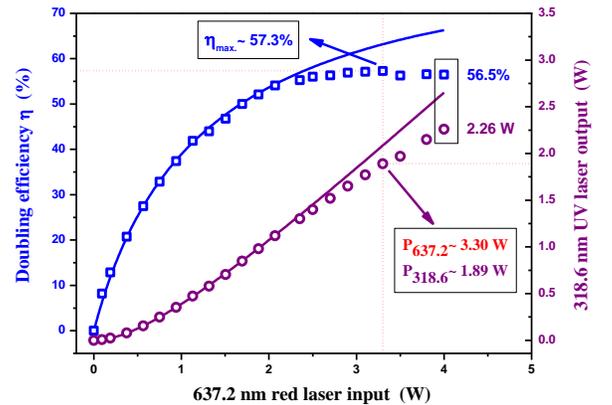

**Fig. 3.** The 318.6-nm UV laser output and the doubling efficiency versus the incident 637.2-nm laser power. Open symbols are the experimental data, while the solid curves are the theoretical results with the parameters $T_1$ = 2.2%, $L$ = 0.67%, and $E_{nl}$ = 6.5×10-5 /W.

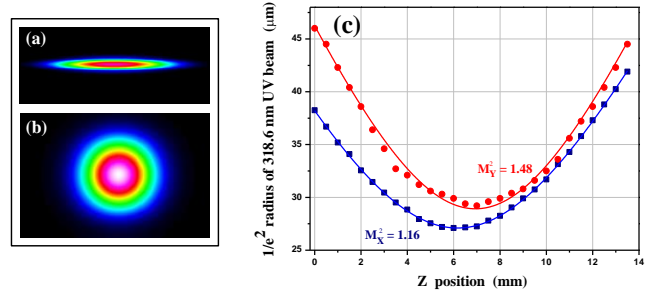

**Fig. 5.** The beam profile and the measured $M^2$ values for the 318.6-nm laser output. On the left are the UV beam intensity profiles (a) before and (b) after shaping. On the right, the measured beam quality factors $M_X^2$ (blue squares) and $M_Y^2$ (red dots).

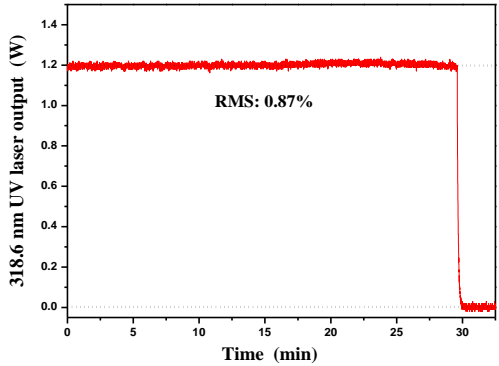

**Fig. 4.** Power stability of the 318.6-nm UV laser output at 1.2 W over 30 min. The typical RMS fluctuation is less than 0.87%.

We monitor the output power of the UV laser with a detector (Thorlabs model PDA25K) after carefully optimizing the parameters of the cavity locking loop, and a typical result is shown in Fig. 4. The root-mean-square (RMS) fluctuation over 30 minutes at 1.2 W is less than 0.87%. This is strong evidence that our locking scheme is robust. To improve the thermal stability, the BBO crystal external temperature is controlled at 19 °C. The fluctuation observed is mainly caused by slow changes in polarization in the fiber amplifiers. These changes are likely attribute to temperature variation and laboratory air disturbances.

The large birefringent walk-off angle (80 mrad) [16] and the bow tie cavity make the output UV beam astigmatic and elliptical. The observed beam profile of 318.6 nm is an oblate ellipse shown in Fig. 5(a) as measured by a slit beam profiler (Thorlabs model BP209-VIS). The beam waist after exiting the cavity is horizontally 2.1 mm and vertically 0.3 mm. To reshape the UV beam, a 200-mm focal length convex lens collimates the horizontal (X) direction, followed 47 mm downstream by a 100-mm focal length cylindrical lens to collimate the vertical (Y) direction. Finally, by using an anamorphic prism pair, we obtain a relatively circular beam with a 0.9-mm X by 0.8-mm Y waist, shown in Fig. 5(b). The beam quality factor $M^2$ is also evaluated in the two orthogonal transverse directions X and Y. As can be seen in Fig. 5(c), $M_X^2$ is 1.16 and $M_Y^2$ is 1.48. The larger $M_Y^2$ is caused by the crystal walk-off.

Considering potential spectroscopic applications, we investigate SHG frequency tunability. A 500-MHz free spectral range (FSR) confocal UV Fabry–Pérot cavity analyzes a frequency sweep of the system. While the 1560.5-nm laser frequency is locked, locking the ring cavity to the 637.2-nm laser, the 1076.9-nm laser is swept slowly and a fringe pattern is generated, shown in Fig. 6. The UV laser can be smoothly tuned across more than 12 FSRs, indicating a continuously tunable range of more than 6 GHz. The typical 1076.9-nm laser sweep speed of 36 GHz/s is a stable slew rate for the ring cavity while locked. Combining the temperature coarse tuning and the PZT fine tuning of the two fiber lasers, we can access a number of cesium atom transitions to Rydberg levels. The free-running frequency fluctuation of the SFG output at 637.2 nm is approximately 22 MHz over 30 min with the 1560.5-nm laser locked to the ULE cavity. In subsequent experiments, we intend to actively stabilize the UV by locking the 1076.9-nm fiber laser to a Rydberg transition using a room temperature cesium vapor cell as reported in reference 12.

To carry out the single-photon Rydberg excitation in cold cesium, the 318.6-nm laser linewidth must be narrow. For example, the cesium $84P_{3/2}$ state has a typical lifetime of 270 μs [26], hence the natural linewidth is estimated to be approximately $2\pi \times 590$ Hz. Our narrow linewidth DFB fiber lasers allow a satisfactory match. Given linewidths of 600 Hz at 1560.5 nm and 2 kHz at 1076.9 nm, we estimate the linewidth of the 318.6-nm UV laser should be less than 10 kHz, which is suitable for driving the single-photon $6S_{1/2}$-nP (n = 70 to 100) Rydberg excitation of cesium atoms.

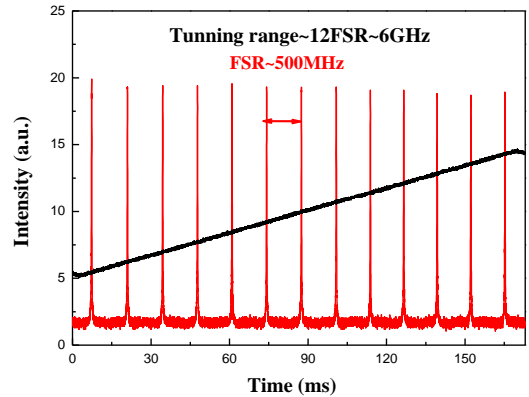

**Fig. 6.** Frequency slew of the 318.6-nm laser monitored by a confocal UV Fabry–Pérot cavity as the 1076.9-nm fiber laser is swept. The tuning range is greater than 6 GHz.

## 4. CONCLUSION

We demonstrate a high-power, narrow-linewidth, and continuously tunable UV laser source at 318.6 nm. It utilizes SFG of infrared fiber lasers followed by SHG in an optimized bow tie ring cavity with a 10 mm Brewster-cut BBO crystal. UV laser output of 2.26 W is obtained with 4.0 W incident at 637.2 nm. The maximum doubling efficiency is 57.3%. UV power fluctuation is 0.87% (RMS) at 1.2 W over 30 min. The continuously tunable range is more than 6 GHz, and the linewidth is estimated to be less than 10 kHz. Our cavity locking technique has advantages in stably locking and slewing such high-power nonlinear laser systems. The Watt-level UV laser will be used in single-step $6S_{1/2}$-nP (n = 70 to 100) Rydberg excitation of cesium atoms. Other potential applications include cold atomic physics [8, 10-14, 16-17, 27], laser material processing, laser medicine, and optical information processing.

**Funding.** National Natural Science Foundation of China (NSFC) (61227902, 61475091, and 11274213); National Major Scientific Research Program of China (2012CB921601).